\documentclass{appolb}
\usepackage{graphicx}
\usepackage{bm}
\usepackage{amsmath}

\begin{document}
\title{Bound-free pair production mechanism in Pb-p collisions at LHC%
}
\author{Melek YILMAZ ~\c{S}ENG\"{u}L
\address{Hali\c{c} University, Engineering Faculty, Electrical and Electronics Engineering, 
             34060, \.{I}stanbul,Turkey}
\\[3mm]
\address{melekaurora@yahoo.com,melekyilmazsengul@halic.edu.tr }
\\[3mm]
}
\maketitle
\begin{abstract}
In this work, cross section calculations of bound-free pair production (BFPP) are done for the mechanism in Pb-p collisions at LHC. BFPP cross section for the asymmetric collisions of Pb-p at the center of mass energies of $\sqrt{s_{NN}}=5,02 TeV$ and $\sqrt{s_{NN}}=8,16 TeV$ is computed. In order to reach the exact results, Monte Carlo integration techniques are utilized to calculate the lowest-order Feynman diagrams amplitudes via the lowest order perturbation theory. Also, in this work our cross section results for BFPP mechanism in Pb-p collisions at LHC are compared with BFPP cross section results obtained in literature, which are reached for Pb-p collisions by using a simple scaling applied to scale BFPP cross section results in Pb-Pb collisions at LHC.
\end{abstract}
  
\section{Introduction}
The main construction aim of the Large Hadron Collider (LHC) is to examine the Quark-Gluon Plasma (QGP) that is only created at high temperatures and densities by the collisions of fully striped lead ($^{208}Pb^{82+}$) ions. In the LHC, heavy ion collisions are the most important case, while the other essential part of the LHC is the proton-proton collisions to discover the Higgs boson. To search the features of QGP, the results of proton-Pb and deuteron-Pb collisions are preferred \cite{1,3}.

In the initial design of LHC, p-Pb collison was not planned, p-Pb collisons have been tried in LHC up to the years 2011/2012 and successfully obtained \cite{4}. The p-Pb experiments that were done at LHC in 2016 are the most successful ones. These experiments were done for two different beam energies and an inverted beam directions. The key parameter of nucleon-nucleon collisions is the center of mass energy $\sqrt{s_{NN}}$. The first part of the p-Pb experiments that were done in 2016 at the center of mass energy $\sqrt{s_{NN}}=5,02 TeV$ .The second part of these experiments was done also at the center of mass energy $\sqrt{s_{NN}}=8,16  TeV$\cite{1}.

During Run 2 (2015-2018), the LHC worked approximately with two times higher energy and achieved Pb–Pb collisions with an order of magnitude higher luminosity compared to Run 1. As a result, the power via BFPP process increased almost by a factor of 20 in comparison with the power of the secondary beams emitted from the interaction points\cite{5}. By this way, the importance of BFPP cross section calculations has increased for the Pb-p collisions. Because of this reason, we concentrate on the asymmetric collisions of Pb-p with accelerating and colliding protons with Pb which was examined for the first time at LHC \cite{4,6,7}. 

The upgrade of the LHC is the High-Luminosity Large Hadron Collider (HL-LHC) to accomplish much more integrated luminosities soon.In
2028, the performance of the HL-LHC p–Pb will mostly be obtained in LHC
Run 3. The p-Pb experiments and calculations will be complementary for the Pb-Pb results \cite{1}. One of the most important impression in heavy ion collider physics is the bound-free pair production to the total cross section and the effects of photonuclear processes. At high energies, these processes highly contribute to the total cross section in symmetric collisions of heavy ions of high charge, such as Pb-Pb collisions. However, the contribution of asymmetric collisions, such as p-Pb collisions cannot be ignored \cite{1}. Before the experiments are to be done, obtaining the confidential cross section calculation results is very important. In this work, for the first time p-Pb cross section calculations are done by the previously tested and the working method. 

\section{Formalism}

\begin{figure}[htb]
\centerline{%
\includegraphics[width=0.47\textwidth]{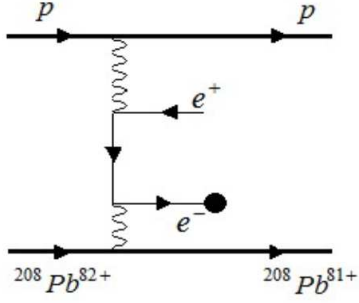}}
\caption{Lowest-order Feynman diagram (direct diagram) for bound-free electron-positron pair production in the asymmetric Pb-p collision \cite{8}}
\label{f}
\end{figure}

In this paper, cross section of BFPP in the asymmetric Pb-p collision as depicted in Fig.~\ref{f} is computed. Monte Carlo method and the semi-classical approximation by utilizing the lowest order perturbation theory in the framework of QED is used to obtain the exact results. Monte Carlo techniques were utilized in the computations of BFPP cross section, and to ensure sufficient convergence of our theoretical results, the integrands were tested at 10 M randomly chosen "positions". In the computations, total numerical error is found to be less than 5~\%{} \cite{8}.

The number of events that creates the quark-gluon plasma depends on the inelastic hadronic cross section in heavy ion collisions. The collision is defined as ultra-peripheral, if the impact parameter $b$ in collisions performs $b>2R$  where $R$ is the nuclear electric radius \cite{1}. In heavy ion collisions with the ion colliding peripherally, the strongly Lorentz contracted electromagnetic fields create a virtual photon flux. In ultra-peripheral collisions, the photon flux causes to occur the lepton pair production probability. This effect is not so important in p-p collisions at LHC, but it is very important with the collisions of ions with an atomic number $Z>>1$. One of the colliding ions may catch a small part of the created electrons and it is finalized in a bound state. This situation is defined as BFPP and leads to the change of the ion charge. At photon energies, this process may even come true a bit smaller than $\hbar\omega=2m_{e}$ as the required energy is decreased by the bound state , where $h=2\pi\hbar$  is the Planck's constant, $\omega$  is the photon angular frequency and $m_{e}$  is the electron mass \cite{1,9,10}.

The first order BFPP process for the symmetric Pb-Pb collisions can be written as

\begin{eqnarray}
\label{1}
   ^{208}Pb^{82+}+\:^{208}Pb^{82+} &\rightarrow& ^{208}Pb^{82+}+\:^{208}Pb^{81+}+\:e^{+}.
\end{eqnarray}

The asymmetric BFPP for Pb-p collision type is

\begin{eqnarray}
\label{2}
   ^{208}Pb^{82+}+\:p &\rightarrow& ^{208}Pb^{81+}+\:p+\:e^{+}.
\end{eqnarray}
For Pb-p collisions, BFPP cross section results are orders of magnitudes smaller than for the symmetric Pb-Pb collisions, since proton generates $1/82^{2}$ times smaller photon flux when it is compared with a lead ion \cite{9,11,12}. 

In BFPP cross section calculations of the asymmetric Pb-p collision, the crossed and direct terms are described by Feynman diagrams in lowest order QED. In this process, Sommerfeld-Maue wave function represents the free positron ($\Psi^{(+)}_{q}$) and Darwin wave function represents the captured electron ($\Psi^{(-)}(\vec{r})$ ). The explicit forms of the wave functions can be found in detail in \cite{8}. By using these wave functions, the BFPP cross section of the asymmetric Pb-p collision for the second order perturbation theory can be written as

\begin{eqnarray}\label{3}
\begin{split}
   \sigma_{BFPP} ={} & 
   \int d^2b\sum_{q<0}\left|\left\langle 
   \Psi^{(-)}\left|S\right|\Psi^{(+)}_q\right\rangle\right|^{2}
   =\frac{\left|N_{+}\right|^{2}}{4\beta^2} \,
   \frac{1}{\pi} \left(\frac{Z}{a_H}\right)^3
   \sum_{\sigma_q} \int \frac{d^3qd^2p_\bot}{(2\pi)^5} \\ & \times\left(\mathcal{D}^{(+)}(q: \mathbf{p}_\bot) + 
         \mathcal{D}^{(-)}(q: \mathbf{q}_\bot  - \mathbf{p}_\bot)\right)^{2} \,,
\end{split}
\end{eqnarray}
with

\begin{eqnarray}\label{4}
   \mathcal{D}^{(+)}(q:\mathbf{p}_\bot) & = &
   H(-\mathbf{p}_\bot: \omega_{pr}) \, 
   H(\mathbf{p}_\bot-\mathbf{q}_\bot: \omega_{Pb}) \,
   \mathcal{T}_q(\mathbf{p}_\bot:+\beta),				
\end{eqnarray}
and
\begin{eqnarray}\label{5}
\begin{split}
   \mathcal{D}^{(-)}(q:\mathbf{q}_\bot-\mathbf{p}_\bot) ={} &    H(\mathbf{p}_\bot-\mathbf{q}_\bot: \omega_{Pb}) \,
   H(-\mathbf{p}_\bot: \omega_{pr}) \\ & \times\mathcal{T}_q(\mathbf{q}_\bot-\mathbf{p}_\bot: -\beta) \, .
\end{split}
\end{eqnarray}

Explicit form of the scalar fields associated with proton "$pr$" and ion "$Pb$" can be expressed in momentum space as a function of the corresponding frequencies as

\begin{eqnarray}\label{6}
   H(-\mathbf{p}_\bot:\omega_{pr})=
   \frac{4\pi Ze}{\left(\frac{Z^2}{a^{2}_{H}}\,+\,
                  \frac{\omega^{2}_{pr}}{\gamma^2\beta^2}\,+\,
		  \mathbf{p}^{2}_{\bot}\right)}, 	
\end{eqnarray}
where $\omega_{pr}$ is the frequency for the proton "$pr$",
\begin{eqnarray}\label{7}
   H(\mathbf{p}_\bot-\mathbf{q}_\bot:\omega_{Pb}) & = &
   \frac{4\pi Ze\gamma^2\beta^2}{
         \left(\omega^{2}_{Pb} + \gamma^2\beta^2(\mathbf{p}_\bot -
	       \mathbf{q}_\bot)^2\right)},
\end{eqnarray}

where $\omega_{Pb}$  is the frequency for the lead "$Pb$". The transition amplitude for both "$pr$" and "$Pb$" can be expressed as follows

\begin{eqnarray}\label{8}
   &   & \hspace*{-0.75cm}
   \mathcal{T}_q(\mathbf{p}_\bot:+\beta) 
   \nonumber \\[0.2cm]
   & = & 
   \sum_s \sum_{\sigma_p}
   \frac{1}{\left(E^{(s)}_{p} - \left(\frac{E^{(-)}+E^{(+)}_{q}}{2}\right) 
            -\beta\frac{q_z}{2}\right)} 
   \left[1 + \frac{\bm{\alpha}\cdot\mathbf{p}}{2m}\right]
   \nonumber \\[0.2cm]
   &   & \times
   <{\textbf{u}}\left|{(1-\beta\alpha_z)}\right|{\textbf{u}^{(s)}_{\sigma_p}}>
   <{\textbf{u}^{(s)}_{\sigma_p}}\left|{(1+\beta\alpha_z)}\right|{
        \textbf{u}^{(+)}_{\sigma_q}}>.
\end{eqnarray}
This term represents the relationship between the intermediate photon lines and the outgoing electron-positron lines. Transition amplitude depends explicitly on the velocity of the ions ($\beta$), transverse momentum of the intermediate-state ($\mathbf{p}_\bot$), parallel momentum of the intermediate-state ($p_z$)  and momentum of the positron ($q$). In this expression,$\textbf{u}^{(s)}_{\sigma_p}$ is the spinor part of the intermediate-state \cite{8}. 

We found the expression that represents the BFPP cross section in terms of impact parameter as given below

\begin{eqnarray}\label{9}
\frac{d\sigma_{BFPP}}{db}=\int^{\infty}_{0}dq q b J_{0}(qb){\cal F}(q).
\end{eqnarray}

In this equation, there is a highly oscillatory Bessel function of order zero and the function  ${\cal F}(q)$ is a six-dimensional integral which can be written as

\begin{eqnarray}\label{10}
{\cal F}(q) &=&\frac{\pi}{8\beta^{2}}\left|N_{+}\right|^{2}\frac{1}{\pi}\left(\frac{Z}{a_{H}}\right)^{3}
\sum_{\sigma_{q}}\int^{2\pi}_{0}d\phi_{q}\int\frac{dq_{z}d^{2}Kd^{2}Q}{(2\pi)^{7}} \nonumber \\
&&\times\lbrace H\left[\frac{1}{2}(\mathbf{Q}-\mathbf{q});\omega_{pr}\right]H\left[\mathbf{-K};\omega_{Pb}\right]\mathcal{T}_{q}\left[-\frac{1}{2}(\mathbf{Q}-\mathbf{q});\beta\right]\nonumber \\
&&
+H\left[\frac{1}{2}(\mathbf{Q}-\mathbf{q});\omega_{pr}\right]H\left[\mathbf{-K};\omega_{Pb}\right]\mathcal{T}_{q}\left[\mathbf{K};-\beta\right]\rbrace \nonumber \\
&&\times\lbrace H\left[\frac{1}{2}(\mathbf{Q}+\mathbf{q});\omega_{pr}\right]H\left[\mathbf{-K};\omega_{Pb}\right]\mathcal{T}_{q}\left[-\frac{1}{2}(\mathbf{Q}+\mathbf{q});\beta\right]\nonumber \\
&&
+H\left[\frac{1}{2}(\mathbf{Q}+\mathbf{q});\omega_{pr}\right]H\left[\mathbf{-K};\omega_{Pb}\right]\mathcal{T}_{q}\left[\mathbf{K};-\beta\right]\rbrace.
\label{fq}
\end{eqnarray}

In Eq.~(\ref{10}), ${Z}/{a_{H}}$ is a term coming from the electron wave function; $N_{+}$  is the normalization constant coming from the positron wave function. Here $\mathbf{Q}$ and $\mathbf{K}$ are the new variables being the functions of $\mathbf{p}$ and $\mathbf{q}$. If Eq.~(\ref{10}) is numerically integrated, the following simple ${\cal F}(q)$ expression is obtained for a fixed value of $q$,

\begin{eqnarray}\label{11}
{\cal F}(q)={\cal F}(0)e^{-aq}=\sigma_{BFPP}e^{-aq}.
\end{eqnarray}

In Eq.~(\ref{11}), ${\cal F}(0)$ is equal to the total cross section of BFPP for Pb-p collisions and calculated at $q=0$. The slope of ${\cal F}(q)$ is not dependent on the energies and charges of the heavy ions and it is equal to the constant value $a = 1.35 \lambda_{C}$  in terms of the reduced Compton wavelength of the electron ($\lambda_{C}=\hbar/mc$ )\cite{13}. 
The probability for BFPP in terms of impact parameter is given in Eq.~(\ref{12}) 

\begin{eqnarray}\label{12}
P_{BFPP}(b)=\frac{1}{2\pi b}\frac{d\sigma_{BFPP}}{db} 
& = & \sigma_{BFPP}\frac{a}{2\pi(a^{2}+b^{2})^{3/2}},
\end{eqnarray}

and the cross section can be written as

\begin{eqnarray}\label{13}
\sigma_{BFPP}=\int^{\infty}_{0}\:P_{BFPP}(b)\:d^{2}b=\int^{\infty}_{0}P_{BFPP}(b)2\pi\:b\:db.
\end{eqnarray}

Detailed information on the BFPP cross section calculations can be found in our previous papers \cite{14,15,16,17}.

\section{Results and Discussions}

At energies of $\sqrt{s_{NN}}=5,02 TeV$ and $\sqrt{s_{NN}}=8,16 TeV$, cross section results of BFPP in p-Pb collisions are first published in \cite{1}. In the experiments, at ALICE, the beam direction is reversed to fill a wider rapidity range, i.e., the proton beams which are initially put in Beam 1 (p–Pb) are directed into Beam 2 (Pb-p) and Pb beams are directed from Beam 2 into Beam 1. In the cross section analysis, the total cross section which depends on the beam direction slightly changes.The p-Pb configuration gives a slightly larger value for cross section than the value obtained in \cite{1}. In this work, our BFPP cross section results for Pb-p collisions are compared with the cross section results that are given in \cite{1} as a table.

They did their BFPP cross section calculations for p-Pb system configuration and energy by using a simple scaling that is applied to scale BFPP cross section results in Pb-Pb collisions at LHC. In \cite{1}, both the EMD (Electromagnetic Dissociation) and BFPP cross sections were calculated. The BFPP into the 1st bound state behaves as explained in \cite{9,11,12},

\begin{eqnarray}\label{14}
\sigma_{1s}=Z^{5}_{1}Z^{2}_{2}alog(\gamma_{c}/\gamma_{0}).
\end{eqnarray}

In this equation, index 1 refers to the ion that captures the electron and the index 2 refers to the projectile ion.$\gamma_{0}$ and $a$ values can be obtained in \cite{11}. The $\gamma_{c}$ is the fixed-target Lorentz factor.

The $EMDm$ cross section is expected to scale approximately like the BFPP cross section (where number $m$ represents the produced electron-positron pairs )\cite{18,19}

\begin{eqnarray}\label{15}
\sigma_{EMDm}\propto Z^{2}_{2}log(\gamma_{c}).
\end{eqnarray}

According to the total scaling behaviour of $\sigma_{BFPP/EMD}\propto Z^{2}_{2}log(\gamma_{c})$  (see Eqs.~(\ref{14}) and~(\ref{15})), the scaling factors are calculated for the fixed target Lorentz factors. This scaling is implemented to the different $EMDm$ and $BFPPm$ cross sections in Pb-Pb collisions. These BFPP cross section results in Pb-p collisions at LHC that are reached in \cite{1} with the center of mass energy $\sqrt{s_{NN}}=5,02 TeV$ and the center of mass energy $\sqrt{s_{NN}}=8,16 TeV$ at LHC are equal to $41,3 mbarn$ and $43,7 mbarn$, respectively. The simple scaling method that is explained above is used to calculate Pb-p collisions cross section results. It is not a well-known and not verified method. It is a kind of approximation to predict the experiment results. Because of these reasons, there is a need for more accurate calculations for the given parameters. In contradistinction to \cite{1}, our method is tested before for symmetric A-A collisions for BFPP calculations many times and worked well. Also, we applied our method for asymmetric BFPP cross section calculations, which is used previously to calculate BFPP cross sections for p-Bi and p-Au collisions at NICA collider and we reached satisfactory results \cite{14}.

BFPP cross section results in Pb-p collisions with the center of mass energy $\sqrt{s_{NN}}=5,02 TeV$ and the center of mass energy $\sqrt{s_{NN}}=8,16 TeV$ at LHC are equal to $31,97 mbarn$ and $34,43 mbarn$ in our work, respectively. When we compare our BFPP cross section results for Pb-p collisions with LHC the results that were reached in \cite{1}, it is seen that our cross section results are approximately 20~\%{} lower than the given results in \cite{1}. The method used in \cite{1} is a rough estimation method to have an idea about the future experiment results, it gives the results approximately. However, our method gives more precise results. Because of this reason, two results differ from each other 20~\%{}. Consequently, we have obtained acceptable values by means of Monte Carlo Method used in \cite{14}, as we expect.

\section{Conclusions}

In the present work, BFPP cross section calculations of Pb-p collisions at the LHC are presented. The BFPP cross section results for Pb-p collisions with the center of mass energy $\sqrt{s_{NN}}=5,02 TeV$ and $\sqrt{s_{NN}}=8,16 TeV$ are in the order of magnitude $31- 35  mbarn$ at the LHC. We compared our cross section values with the values given in \cite{1}. By doing these calculations for Pb-p collisions, we wish to give contribution to the experiments that will be performed in LHC. These calculations were done for the first time in \cite{1} by using a simple scaling method. However, in our work, we used Monte Carlo method that is described in detail above and reached the confidential results as obtained in our previous works for different problems utilizing the same method. In this work, we did only BFPP cross section calculations of Pb-p collisions at the LHC. In the future works, we are planning to do EMD cross section calculations of Pb-p collisions at the LHC and compare our results with the future expected experiment results.


\end{document}